\documentclass[conference]{IEEEtran}
\IEEEoverridecommandlockouts
\usepackage{cite}
\usepackage{amsmath,amssymb,amsfonts}
\usepackage{algorithmic}
\usepackage{graphicx}
\usepackage{textcomp}
\usepackage{xcolor}
\def\BibTeX{{\rm B\kern-.05em{\sc i\kern-.025em b}\kern-.08em
    T\kern-.1667em\lower.7ex\hbox{E}\kern-.125emX}}

\begin{document}
\title{Effects of Different Q-swaps Modes on
Percolation Threshold in Small-world
Quantum Networks}

\author{{JianXiong~Liang, Xiaoguang~Chen, Yaoyao~Wang
		}\\
		\IEEEauthorblockA{
			\textit{Department of Communications Science and Engineering}\\
			\textit{School of Information Science and Technology, Fudan University}\\
               \textit{Shanghai, 200433, China}\\
			\textit{(e-mails:  Xiaoguangchen@fudan.edu.cn)}
		}
}

\maketitle

\begin{abstract}
Quantum networks are interconnected by nodes, between singlets which are formed to ensure the successful transmission of information with a probability of 1. However, in real quantum networks, nodes often share a partially entangled state instead of a singlet due to factors such as environmental noise. Therefore, it is necessary to convert the partially entangled state into a singlet for efficient communication. Percolation happens during the conversion of connected edges in the whole network. As a result, when the singlet conversion probability (SCP) is greater than the percolation threshold, a giant interconnected cluster that meets the basic requirements of communication will appear in the network. The percolation threshold of the network reveals the minimum resources required to carry out large-scale quantum communication. In this paper, we investigate the effect of different q-swaps on the percolation threshold in quantum entanglement percolation of small-world networks. We show that Quantum Entanglement Percolation (QEP) has a better percolation performance than Classical Entanglement Percolation (CEP). By using different q-swaps in Watts–Strogatz (WS) small-world networks and Kleinberg networks for simulation, we also show that the percolation threshold is minimized when SCP is equal to the average degree of the network. Furthermore, we introduce quantum walk as a new scheme to have an extra reduction in the percolation threshold.
\end{abstract}


\section{Introduction}
\label{sec:introduction}
Quantum networks lays the foundation of future quantum communication, and with the development of quantum network theory, researches have been conducted on how the maximally entangled states are distributed in quantum networks [1-5]. Nodes in a quantum network are connected by entangled states, most of which are partially entangled ones. Pure partially entangled states can be converted to maximally entangled states by Local operations and Classical Communications (LOCC) so that any two nodes in a quantum network can pass information by maximally entangled states with a probability of 1. In order to communicate more efficiently, it is important to use resources as few as possible to establish maximally entangled states throughout the network while ensuring that the network is capable of basic communications. Another challenge is that maximally entangled states are vulnerable to noise from environments and other sources and thus difficult to maintain for a long time. Maximally entangled state will suffer from decoherence and be converted to a partially entangled state, which can be expressed as [6]
\begin{equation}
	\left| \varphi \right\rangle =\sum\limits_{i=1}^{n}{{{\lambda }_{i}}}\left| ii \right\rangle \quad {{\lambda }_{i}}\ge 0, 
	\label{eq1}
\end{equation}
where ${\lambda }_{i}$ are non-negative Schmidt coefficients. We assume that any two nodes in the quantum network are interconnected by bipartite entangled states, which can be expressed as $\left| \varphi  \right\rangle ={{\lambda }_{1}}\left| 00 \right\rangle +{{\lambda }_{2}}\left| 11 \right\rangle $. $\left| \varphi  \right\rangle$ can be converted to a singlet with a probability $p$, which is called Singlet Conversion Probability (SCP). For a partially entangled state between two nodes, the conversion is carried out with probability $p=\min (1,2(1-{{\lambda }_{1}}))$ [7], and there will be no entanglement between these two nodes if the conversion fails. 

Entanglement Percolation can be used to generate maximally entangled states between any two nodes in the network. There are two types of percolation models: the bond percolation model and the site percolation model. In the bond percolation model, the nodes of the quantum network are fixed and the edges in the quantum network are partially entangled states. The process of percolation starts from any edge with SCP of $p$. If the conversion succeeds, the edge will be occupied; otherwise it is missed. As the edges in the network keep being occupied, a giant interconnected cluster is formed which serves as the real communication network. For the site percolation, on the other hand, it starts from any node in the network and occupies that node with probability $P$ and misses it with probability $1-P$. In both percolation models, clusters start to appear in the network as $P$ increases. A giant cluster will show up when $P>p^*$, where $p^*$ is the percolation threshold that satisfies ${p}^{*}=\frac{1}{{{{{g}'}}_{r}}(1)}$. ${{{{g}'}}_{r}}(1)$ is the probability of bond conversion that depends only on $\left\langle k \right\rangle $ and $\left\langle k^2 \right\rangle $, where $k$ is the value of degree of the node and $\left\langle \cdot \right\rangle $ denotes the mean. This giant cluster has a similar magnitude compared to the entire quantum network and is capable of basic communication. In quantum networks, the percolation threshold of the network implies the minimum resources required to carry out large-scale and long-distance quantum communication, so it is essential to study how to reduce the percolation threshold in small-world quantum networks [9], which is one of the typical networks in real world.

One useful method of reducing percolation threshold is Quantum Entanglement Percolation (QEP). Although entanglement percolation is directly carried out in Classical Entanglement Percolation (CEP), additional quantum pre-processing [10] is needed in QEP: the qubits on repeaters are measured to implement entanglement swapping, which connects two disconnected nodes. This kind of pre-processing is called q-swap, which changes the structure and the percolation threshold of the original network.

In order to modify the network structure and further improve the percolation threshold, we also use coined quantum walk. The process of quantum walk can be seen as specific operations that implement certain functions [11-12]. And the former research work has shown that multi-coin quantum walk can be used to generate entangled states between targeted qubits [13]. Therefore, it can also serve as pre-processing and form new entanglements in the network, which meets our requirements.

The paper is organized as follows. First, we will briefly introduce different q-swaps in one-dimensional and two-dimensional networks. Next, we will simulate and analyze the effects of different q-swaps on the percolation threshold in one-dimensional and two-dimensional networks respectively. We will also investigate the effect of adding randomly connected edges on the percolation threshold of the whole network. Besides, since the Kleinberg network is directed, we look into the outgoing and incoming degrees of the nodes separately when carrying out the q-swap operation and compare the difference between the Kleinberg network and the square lattice in terms of the percolation threshold. Then, we will introduce quantum walk as a new scheme and compare its effects to those of q-swap. Finally, appropriate values of q were obtained.

\section{Entanglement Percolation in Small-world Networks}\label{Section2}
Most of the real-world networks shares some similar characteristics to small-world networks, so is the structure of quantum network [14]. Therefore, it's reasonable to study how q-swap affects the percolation threshold in the process of quantum percolation in small-world networks. There are two common entanglement percolation methods: CEP and QEP. Compared to normal entanglement percolation in CEP, q-swap should be carried out first before entanglement percolation in QEP to change the network. Here, we introduce how q-swap will change the structure of one-dimensional and two-dimensional small-world networks and thus affecting the percolation threshold of the corresponding networks.

\subsection{One-dimensional Networks}
Watts–Strogatz (WS) network is constructed by disconnecting and reconnecting edges with probability $p$ starting from the nearest-neighbor coupled network [15]. As one of the typical small-world networks, WS network has a higher clustering coefficient and a lower average distance compared to normal networks. In quantum networks, we assume that all the connected edges in a WS network are in state $\left| \varphi  \right\rangle$. To analyze how the percolation threshold is affected by q-swap, we look into a WS network with an average degree of 6.

\begin{figure}[htbp]
\centerline{\includegraphics[width=1\linewidth]{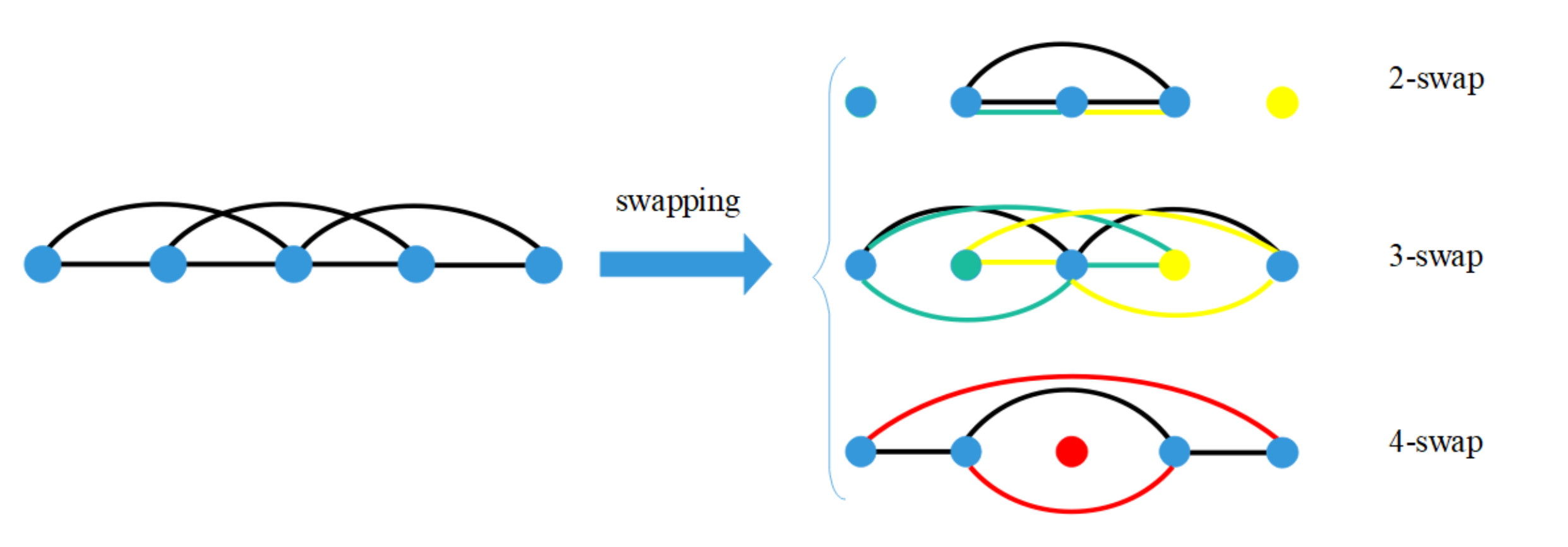}}
\caption{Effects of different q-swaps on the structures of one-dimensional quantum network.}
\label{fig1}
\end{figure}

As shown in Fig.~\ref{fig1}, for a one-dimensional quantum network, different network structures are obtained when $q$ is equal to 2, 3, and 4 respectively and the degree value of each node also changes. As a result, the percolation threshold also changes since $p^*$ is relative to degree $k$. The results of different q-swaps also vary. For example, if a new connected edge is generated between two disconnected nodes through entanglement swapping, it will exhibit as mixed state and the neighboring node that has been measured cannot be involved into other entanglement swapping operations any more. But its SCP is identical to that of the original state $\left| \varphi  \right\rangle$ [16]. So the degree value of the node's neighboring node is affected due to entanglement swapping.

\subsection{Two-dimensional Networks}
\begin{figure}[htbp]
\centerline{\includegraphics[width=0.8\linewidth]{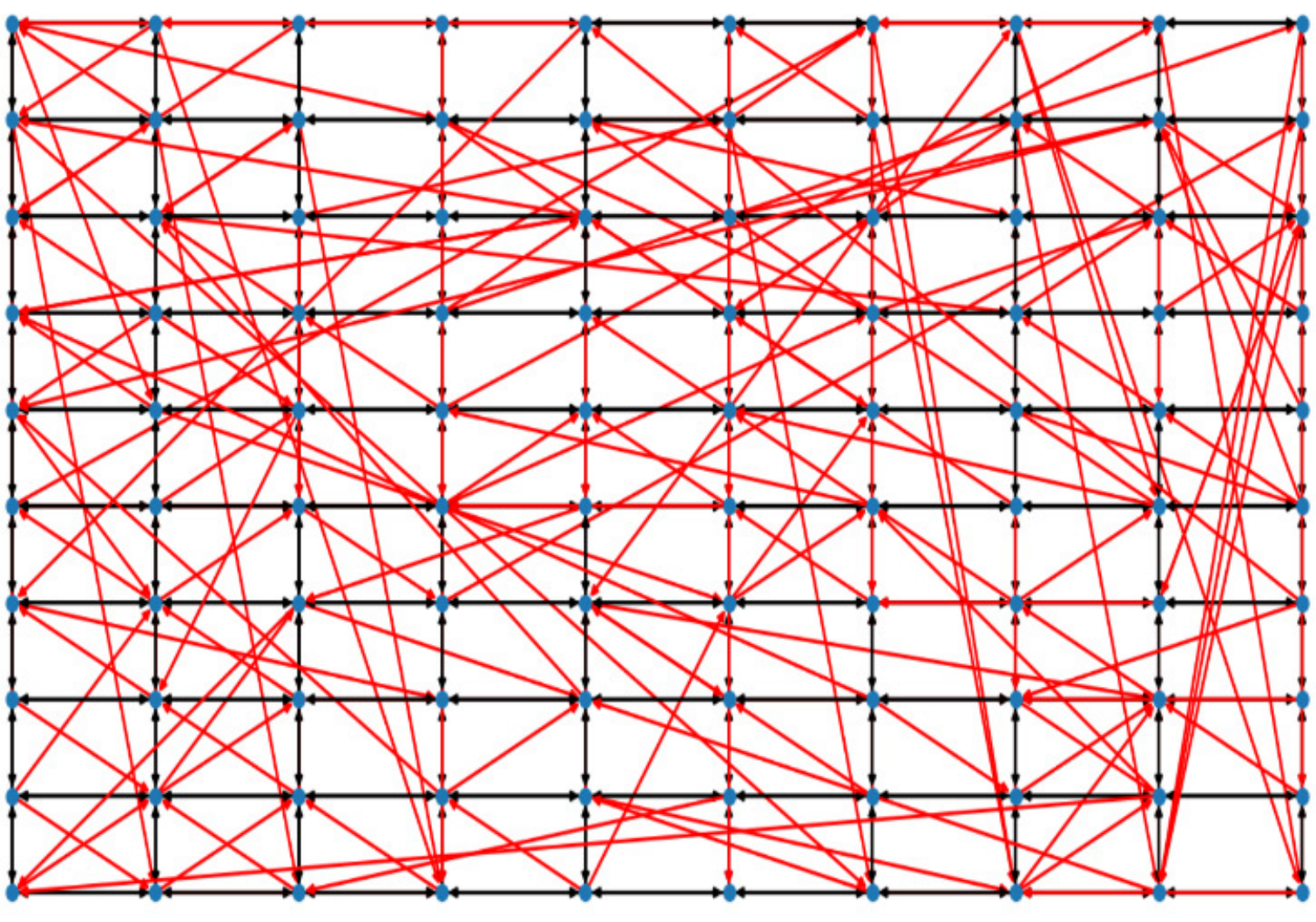}}
\caption{A two-dimensional Kleinberg small-world network (long-range connections are drawn in red and short-range connections are drawn in black; each blue dot indicates a quantum memory).}
\label{fig2}
\end{figure}

Here, we take the two-dimensional Kleinberg small-world network as an example. Fig.~\ref{fig2} shows a Kleinberg network. The two-dimensional Kleinberg small-world network is constructed as follows [17]: a two-dimensional square lattice is used as the underlying network, in which each node in the network establishes bidirectional edges with its four neighboring nodes. To ensure the small-world characteristics, after establishing the underlying network, extra long-range edges are established. Each node has $z$ directed long-range edges pointing to the other nodes in the network, and the probability that a long-range edge exists between any two nodes $x(m,n)$ and $y(i,j)$ in the network is [18]
\begin{equation}
	p(x,y)=\frac{d{{(x,y)}^{-g}}}{\sum\limits_{x\ne y}{d}{{(x,y)}^{-g}}},
	\label{eq2}
\end{equation}
where $d(x,y)=\|m-i\|+\|n-j\|$ and $g$ is the clustering coefficient whose only optimal solution is $g=2$. We also assume that all the connected edges in Kleinberg small-world network are in state $\left| \varphi  \right\rangle$.

\begin{figure}[htbp]
\centerline{\includegraphics[width=0.8\linewidth]{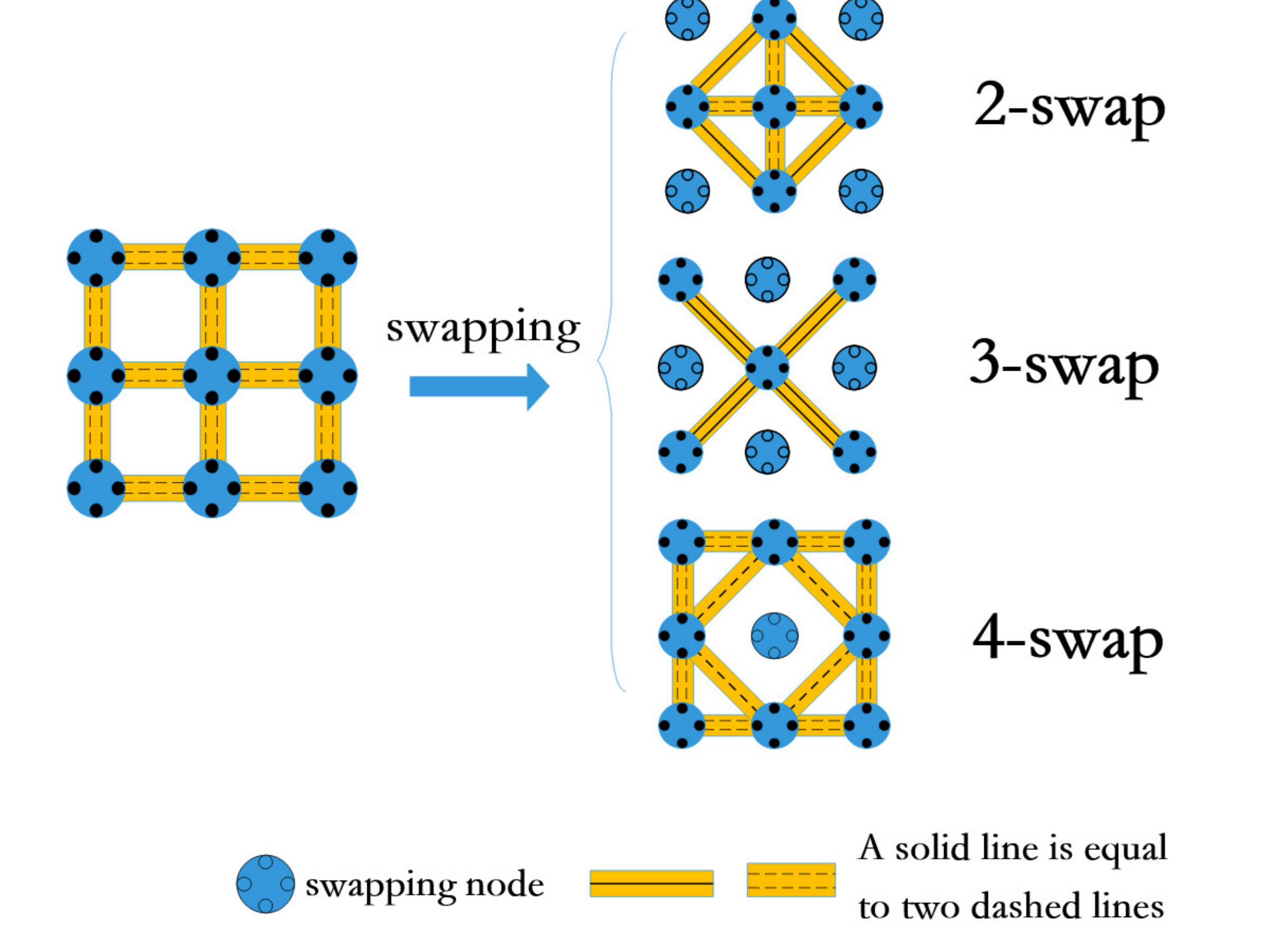}}
\caption{Effects of different q on the structure of the underlying networks of a Kleinberg network.}
\label{fig3}
\end{figure}

Similar to one-dimensional networks, different q-swap operations will result in different two-dimensional network structures. Different structures are obtained after implementing different q-swap operations to a Kleinberg small-world quantum network when $q$ is equal to 2, 3, and 4, as shown in Fig.~\ref{fig3}. Each structure has a different percolation threshold as well. To study the effect of entanglement swapping on the percolation threshold of the network for nodes of different degrees, we carry out CEP and QEP simulations in square lattice networks and Kleinberg small-world quantum networks, respectively. Unlike square lattice, the connected edges of Kleinberg networks are directed ones, so entering edges and exiting edges need to be considered separately when pre-processing the network.

\begin{figure}[htbp]
\centerline{\includegraphics[width=0.8\linewidth]{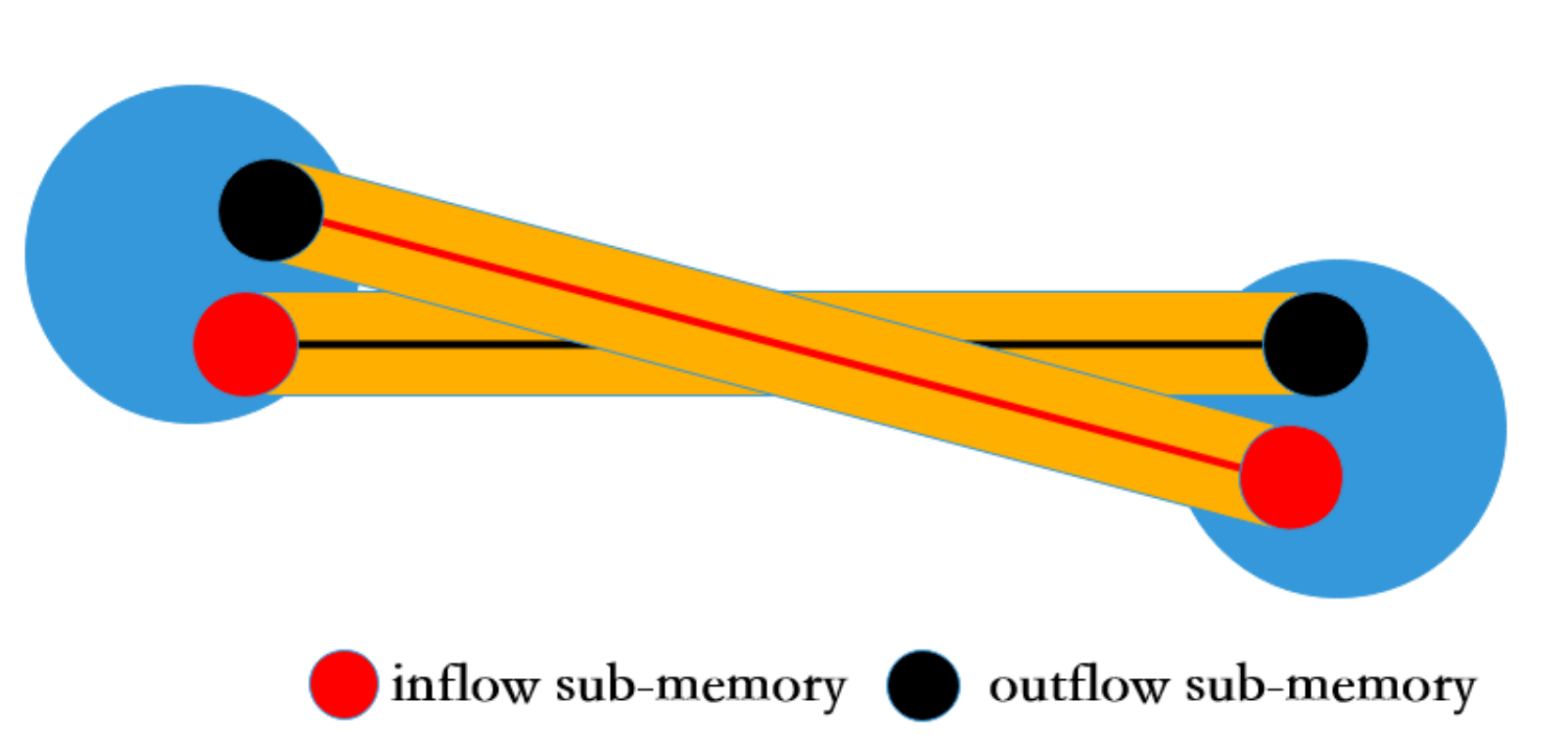}}
\caption{In-flow and out-flow memory in nodes.}
\label{fig4}
\end{figure}

As shown in Fig.~\ref{fig4}, there are two kinds of memories (in-flow and out-flow) in the nodes of the Kleinberg small-world quantum network, and entanglement swapping can only be implemented between connected edges of memories with the same in-flow or out-flow directions. The QEP simulation is performed as follows: nodes with an in-degree of 5 and an out-degree of 6 are selected for the q-swap operation respectively.

\section{Entanglement Percolation in Small-world Networks}\label{Section3}
In this part, we simulate and analyze the WS small-world network and the Kleinberg small-world quantum network respectively.

\begin{figure}[htbp]
\centerline{\includegraphics[width=0.8\linewidth]{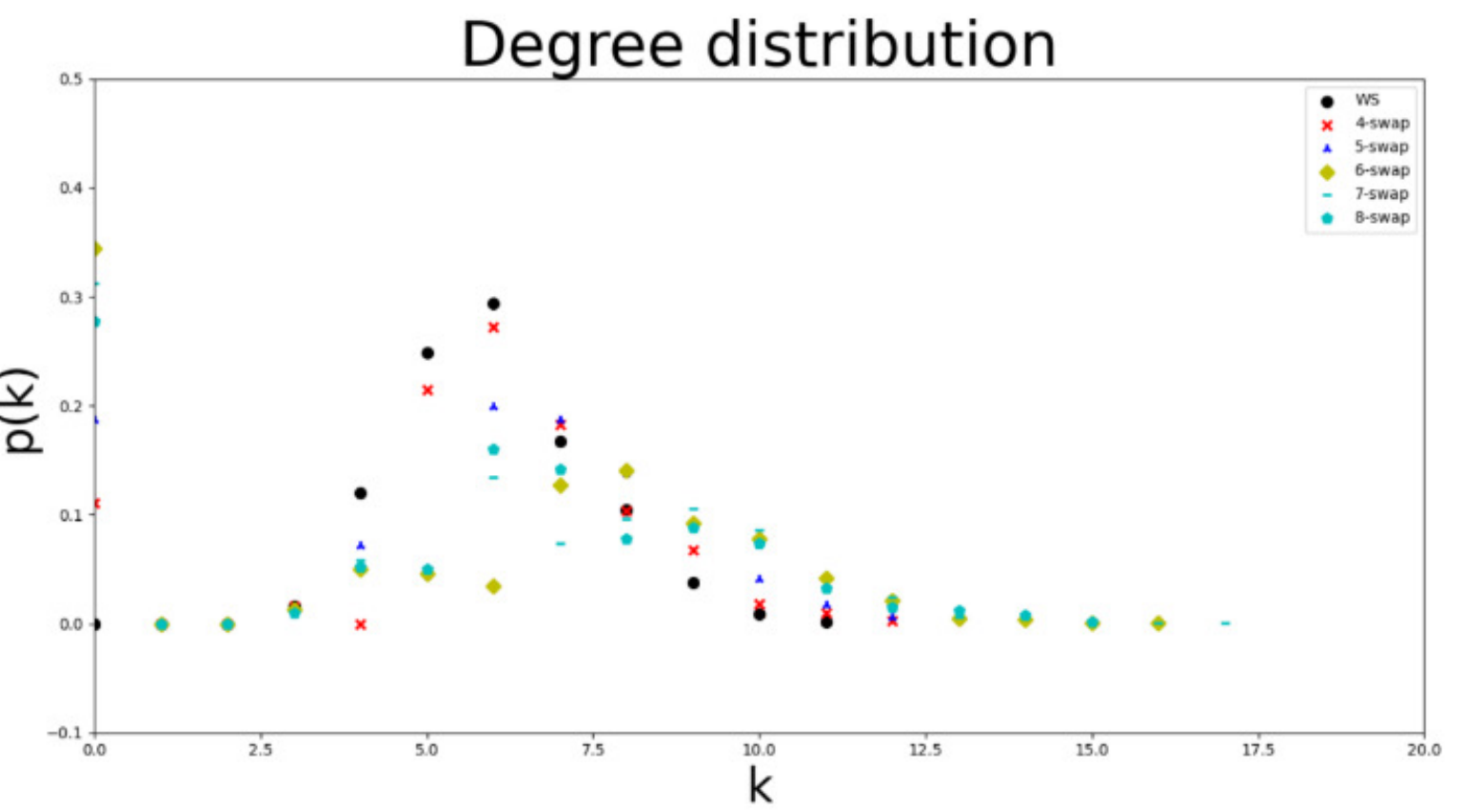}}
\caption{Degree distribution of WS network and networks after quantum pre-processing.}
\label{fig5}
\end{figure}

\begin{figure}[htbp]
\centerline{\includegraphics[width=0.8\linewidth]{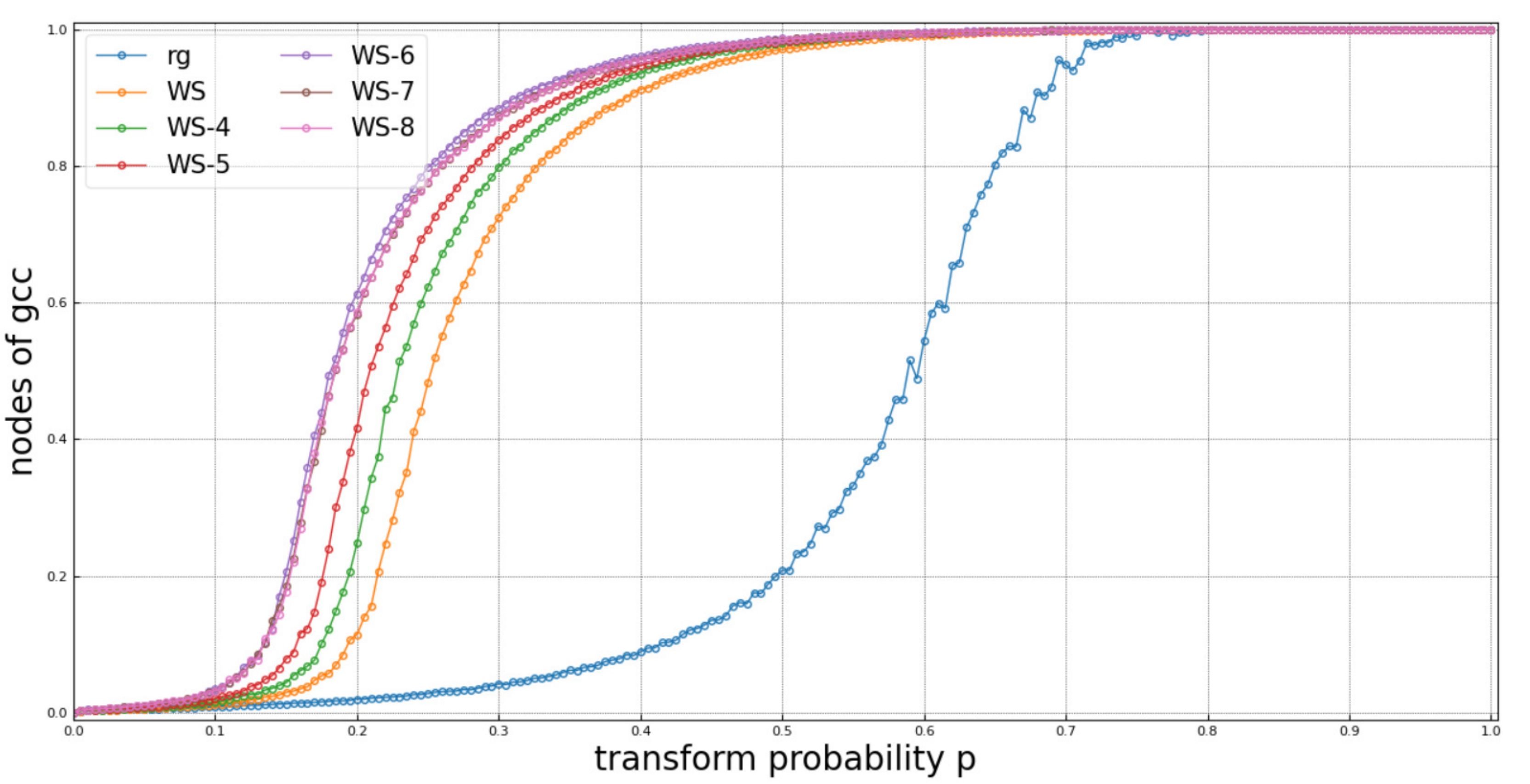}}
\caption{Relationships between GCC and SCP in a WS network with an average degree of 6.}
\label{fig6}
\end{figure}

\subsection{One-dimensional Networks}
To study the effect of q-swap on the percolation threshold of WS small-world networks, we select a WS quantum small-world network with an average network degree of 6 and 1000 nodes. CEP and QEP simulations are implemented on the WS quantum small-world network, respectively. In QEP pre-processing, $q$ is equal to 4, 5, 6, 7, and 8, respectively. The degree distribution of the networks after pre-processing is shown in Fig.~\ref{fig5}. The values of $\left\langle k \right\rangle $ and $\left\langle k^2 \right\rangle $ for each network are $(6, 37.988)$, $(5.856, 39.566)$, $(5.708, 42.598)$, $(5.302, 45.436)$, $(5.246, 44.2)$, and $(5.398, 44.068)$, respectively. The q-swap operation causes intermediate nodes to lose connectivity with neighboring nodes and results in a lower average network degree compared to the original WS network. However, the network degree distribution is more uniform after q-swap and the phenomena become more obvious as the value of $q$ increases. Meanwhile, we can also see that the number of nodes with smaller degree values increases and nodes with larger degree values start to show up as $q$ increases.

In a small-world quantum network, nodes with large degree values have a more significant impact on $\left\langle k^2 \right\rangle $ than those with small degree values. Since the percolation threshold is only related to $\left\langle k \right\rangle $ and $\left\langle k^2 \right\rangle $, we only take into account the degree distribution of the network in the analysis.

The relations between the scale of Giant Connected Component (GCC) and the transform probability, namely SCP, in a 6-degree WS network and networks after different q-swap operations are shown in Fig.~\ref{fig6} where GCC stands for the ratio of the number of nodes contained in the largest cluster in the network to the total number of nodes in the network given a certain SCP. The figure also includes the regular one-dimensional network (rg) with an average degree of 6.

We can see from Fig.~\ref{fig6} that the percolation threshold of the WS small-world network is much lower than that of the one-dimensional regular network due to the disconnect-reconnect operation. All the percolation thresholds of networks after q-swap operations are smaller than that of the original WS. The percolation threshold is minimized when $q=6$. Almost all nodes in the network are connected except for the nodes in regular network when $p \geq 0.5$. When $q$ is equal to 6, 7, and 8, the relationship between GCC and SCP are basically the same, so are the percolation thresholds. During percolation, nodes with a larger degree value that show up in 6,7,8-swap networks, are more likely to form a connection successfully. However, the number of these nodes is fairly small, which means that although a lower percolation threshold can be achieved, the changing rate of GCC with SCP decreases after these nodes form connections with others. In order to meet the basic requirements of quantum communication, efforts should be made to make sure that GCC is as large as possible and the percolation threshold is as small as possible to reduce the loss when establishing the network. Therefore, $q=6$ should be the ideal value to achieve the best QEP performance.

\begin{figure}[htbp]
\centerline{\includegraphics[width=0.8\linewidth]{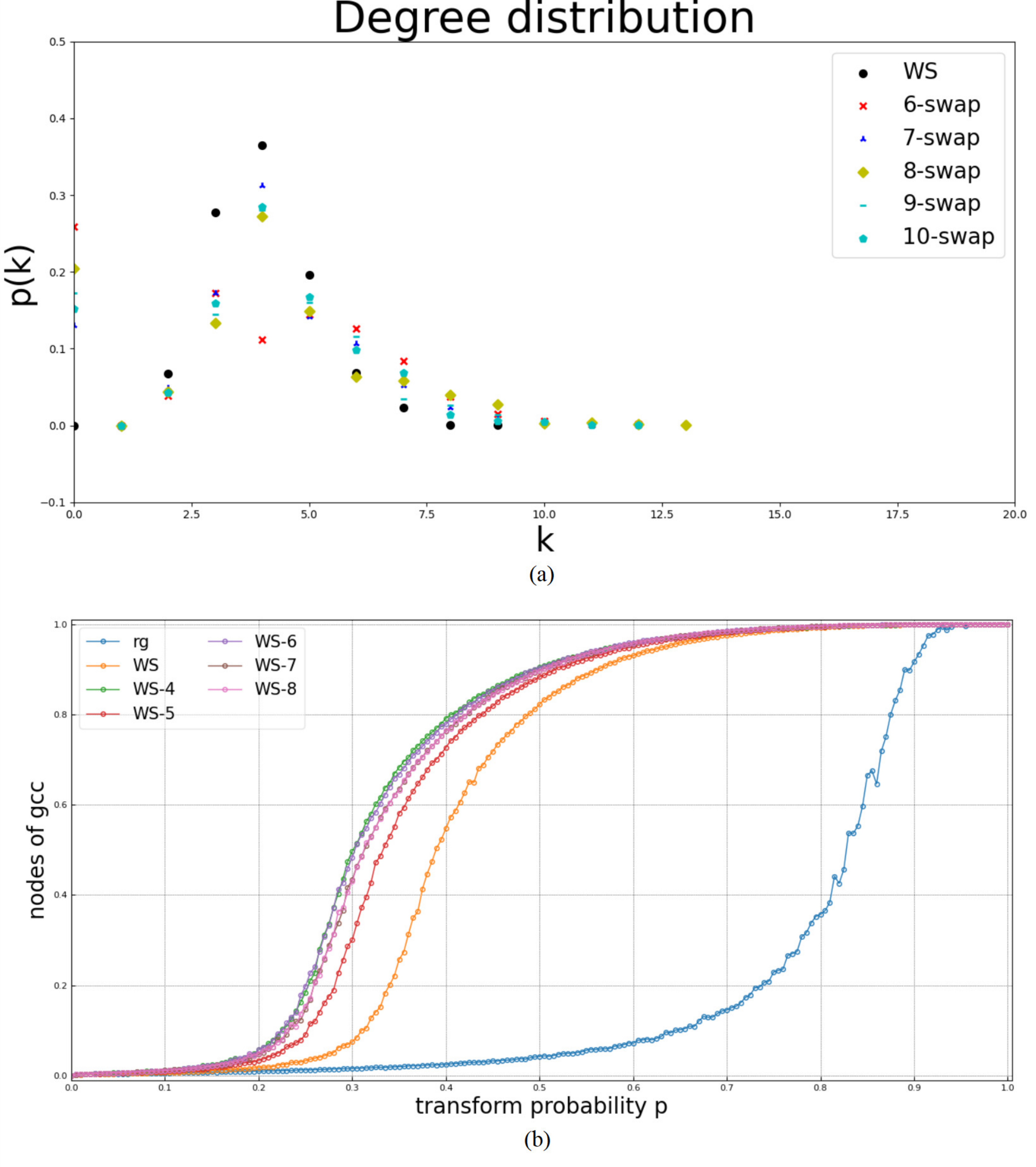}}
\caption{A WS network of an average degree of 4 after q-swap (a) Degree distribution (b)Relationship between GCC and SCP.}
\label{fig7}
\end{figure}

\begin{figure}[htbp]
\centerline{\includegraphics[width=0.8\linewidth]{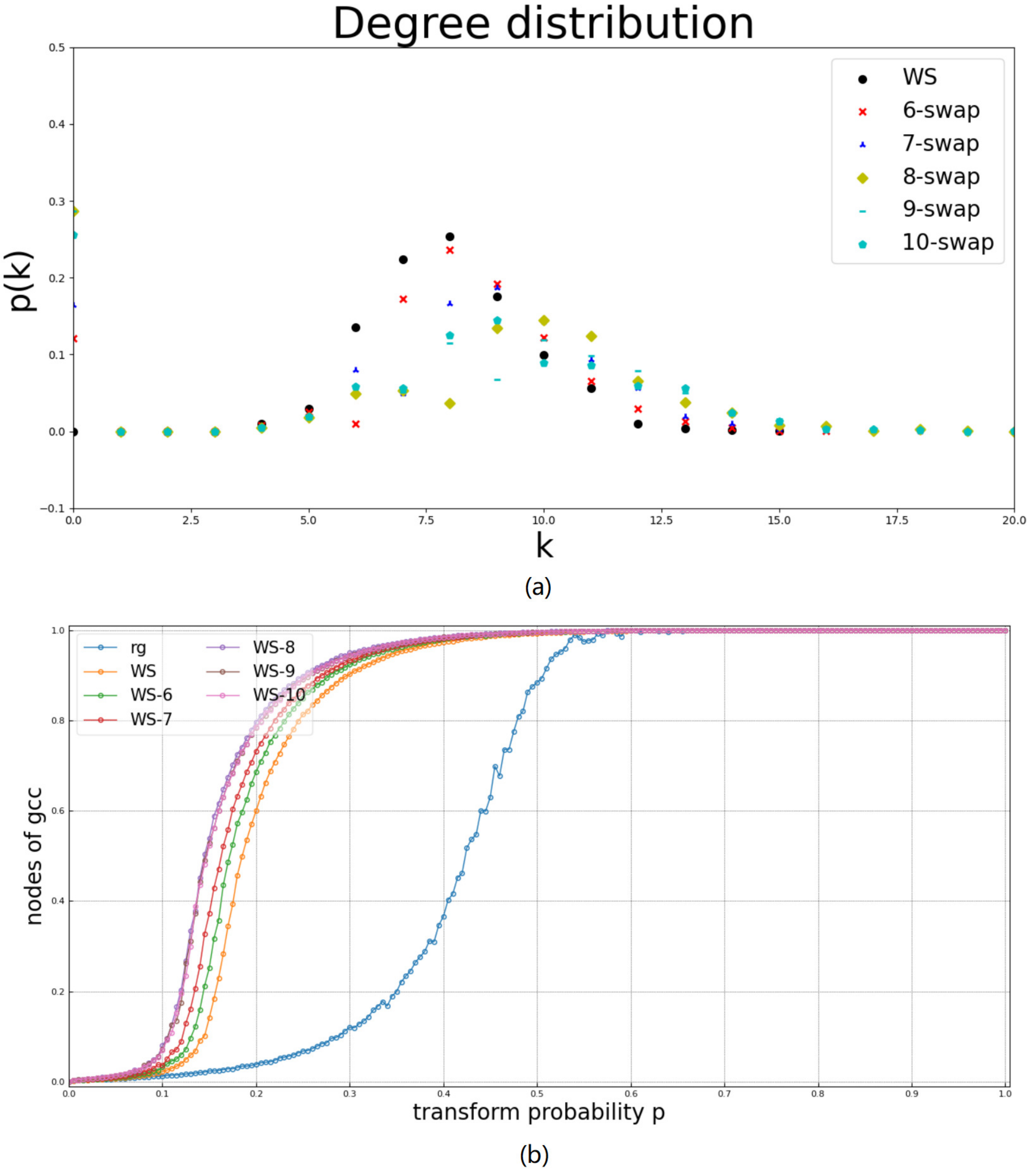}}
\caption{A WS network of an average degree of 8 after q-swap (a) Degree distribution (b)Relationship between GCC and SCP.}
\label{fig8}
\end{figure}

In order to study the effect of the average degree of the network on the percolation threshold of WS small-world networks after q-swap, we implement percolation in WS small-world networks with an average degree of 4 and 8 respectively and analyze the results. For the WS network with an average degree of 4, the value of $q$ ranges from 4 to 8. For the WS network with an average degree of 8, the value of $q$ ranges from 6 to 10. The results, including the network degree distribution and the relationships between GCC and SCP, are shown in Fig.~\ref{fig7} and Fig.~\ref{fig8}. We can see that when the value of in pre-processing is equal to the average degree of the network, we obtain the minimum percolation threshold, no matter what the average degree of the WS network is.

\subsection{Two-dimensional Networks}
\begin{figure}[htbp]
\centerline{\includegraphics[width=0.8\linewidth]{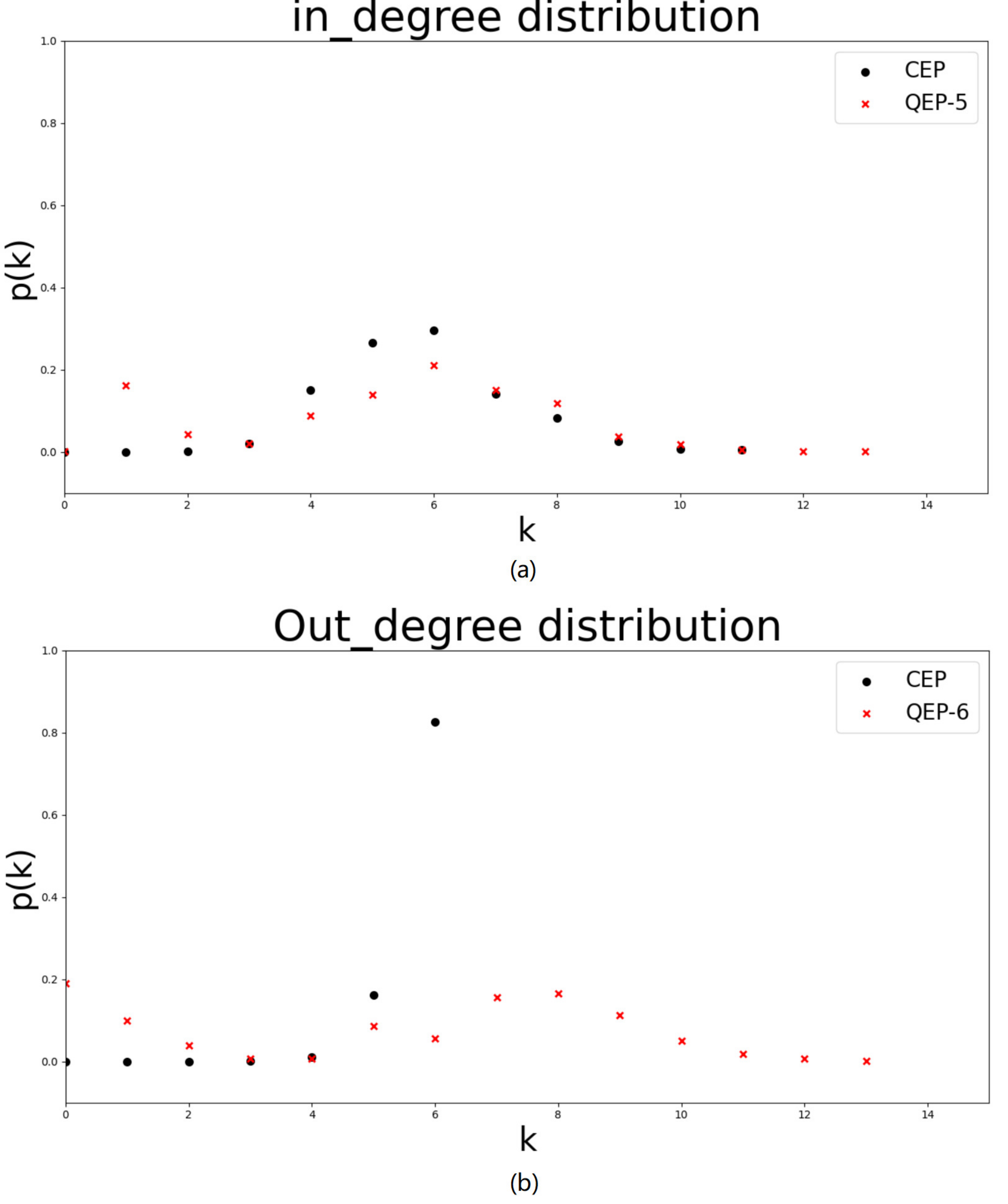}}
\caption{In-degree and out-degree distribution of the Kleinberg network after q-swap (a)In-degree (b)Out-degree.}
\label{fig9}
\end{figure}

Different from one-dimensional WS small-world network, Kleinberg network is a directed one and nodes with an in-degree of $q$ and an out-degree of $q$ should be considered independently when implementing q-swap. The established Kleinberg network has 2 long-ranged edges ($z=2$), a size of $30 \times 30$ and a clustering coefficient of 2 ($g=2$). A $30 \times 30$ square lattice network is selected for comparison so that we can investigate the difference between the percolation threshold of a two-dimensional small-world network and a regular network. The in-degree distribution and the out-degree distribution of Kleinberg network are shown in Fig.~\ref{fig9}. Since all the nodes have an out-degree of 6 except for the ones on the four edges, we chose the node with an in-degree of 5 and the node with an out-degree of 6 for the q-swap operation respectively.

\begin{figure}[htbp]
\centerline{\includegraphics[width=0.8\linewidth]{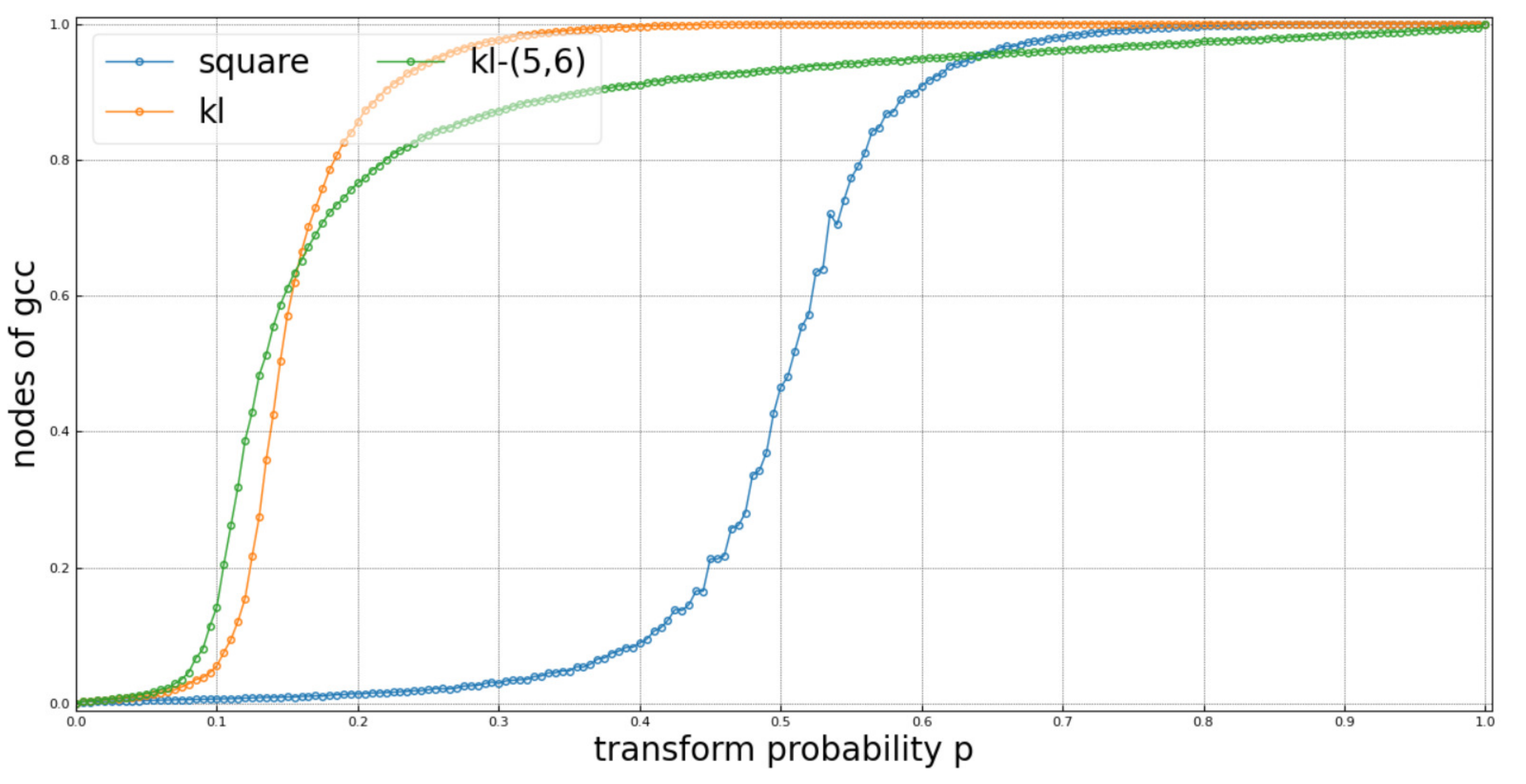}}
\caption{Relationships between GCC and SCP in Kleinberg small-world network and square lattice.}
\label{fig10}
\end{figure}

If two nodes are connected in both directions, the probability of successful conversion of the connected edges between these two nodes is twice as high as that of the unidirectional connected edges. The result of the simulation is shown in Fig.~\ref{fig10} and we can see that the percolation threshold is significantly lower in Kleinberg networks. The q-swap operation can lower the percolation threshold even further. However, the cluster is formed more slowly, which is similar to the case in one-dimensional networks.

\begin{figure}[htbp]
\centerline{\includegraphics[width=1.1\linewidth]{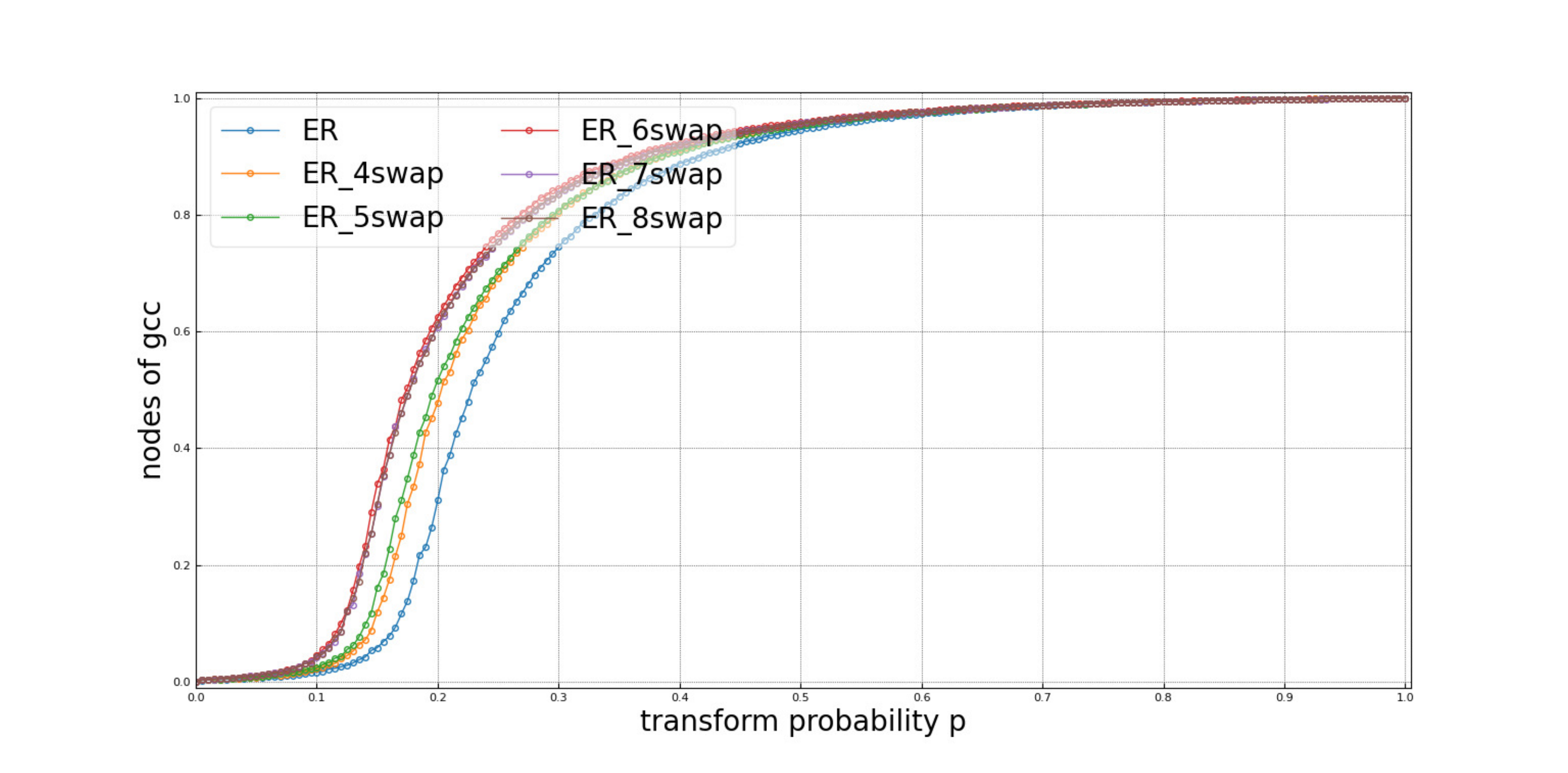}}
\caption{Relationships between GCC and SCP in an ER network.}
\label{fig11}
\end{figure}

Meanwhile, we also analyzed the results in an Erdős–Rényi (ER) network with the same number of nodes $N=1000$ and the same average network degree $\left\langle k \right\rangle =6$, which is shown in Fig.~\ref{fig11}. We can see that the percolation threshold is minimized when 6-swap is performed, which is identical to the case in a WS network: when doing q-swap operations, the percolation threshold is minimized if $q =\left\langle k \right\rangle$.

\section{Percolation in One-dimensional Network using Quantum Walk}\label{Section4}
In order to change the structure of the network and further decrease the percolation threshold, we introduce the quantum walk scheme. We can use multi-coin quantum walk to generate entangled states of multiple qubits, which forms extra connections between nodes.

\begin{figure}[htbp]
\centerline{\includegraphics[width=0.8\linewidth]{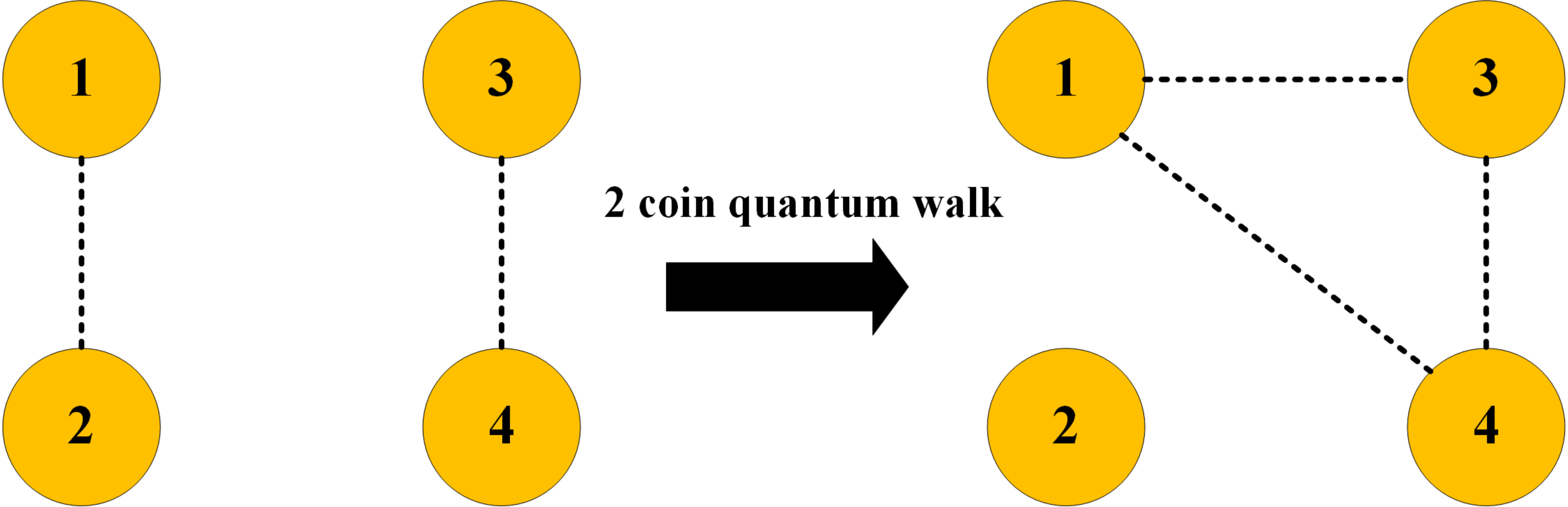}}
\caption{Generate new entanglement connection using quantum walk. The circles indicate qubits and dashed lines indicate entanglement. Entanglement between 1,3 and 4 can be achieved.}
\label{fig12}
\end{figure}

The space of quantum walk can be described as $H_P\otimes H_C$ where $H_P$ is the position space expanded by position state $\left\{ \left| x \right\rangle ,x\in Z \right\}$ and $H_C$ is the coin space expanded by two orthogonal coin states $\left\{ \left| \uparrow  \right\rangle ,\left| \downarrow  \right\rangle  \right\}$. The corresponding unitary operator of each step of quantum walk can be described as [13] 
\begin{equation}
	U=S \cdot \left( I_P\otimes H_C \right),
	\label{eq3}
\end{equation}
where $S$ is the shift operator that determines the movement of the particle and $I_P$ is the identity operator that acts on position space $H_P$. $H_C$ is the coin operator that indicates the internal degree of freedom of the particle. In a more general situation, $H_C$ indicates the possible direction of the movements.

In a graph with $N(N \geq 2)$ nodes, we carry out multi-coin quantum walk. Assume that the initial state of the system is $\left| \Psi \left( 0 \right) \right\rangle $, the system state after $k$ steps is 
\begin{equation}
	\left| \Psi \left( k \right) \right\rangle =\left( {{U}_{k}}{{U}_{k-1}}\cdots {{U}_{1}} \right)\left| \Psi \left( 0 \right) \right\rangle .
	\label{eq4}
\end{equation}

We only consider a complete graph and shift operator $S^{n-com}$ can be expressed as[19]
\begin{equation}
	S^{n-com}=\sum\limits_{x,i=0}^{n-1}{\left| \left( x+i \right)\bmod n \right\rangle \left\langle  x \right|\otimes \left| i \right\rangle \left\langle  i \right|},
	\label{eq5}
\end{equation}
and when $m=2$, the corresponding shift operator $S^{2-com}$ can be expressed as
\begin{equation}
	\begin{aligned}
		{{S}^{2-com}}= &\left( \left| 0 \right\rangle \left\langle  0 \right|\otimes \left| 1 \right\rangle \left\langle  1 \right| \right)\otimes \left| 0 \right\rangle \left\langle  0 \right| \\ 
		& +\left( \left| 1 \right\rangle \left\langle  0 \right|\otimes \left| 0 \right\rangle \left\langle  1 \right| \right)\otimes \left| 1 \right\rangle \left\langle  1 \right|. \\ 
	\end{aligned}
	\label{eq6}
\end{equation}

Therefore, for a system with two Bell states, we can generate a new entanglement connection between the existing ones, as shown in Fig.~\ref{fig12}. The newly generated state is actually GHZ state which satisfies our requirement of having extra connections between qubits. The detailed process is described as follows.

\begin{figure}[htbp]
\centerline{\includegraphics[width=0.8\linewidth]{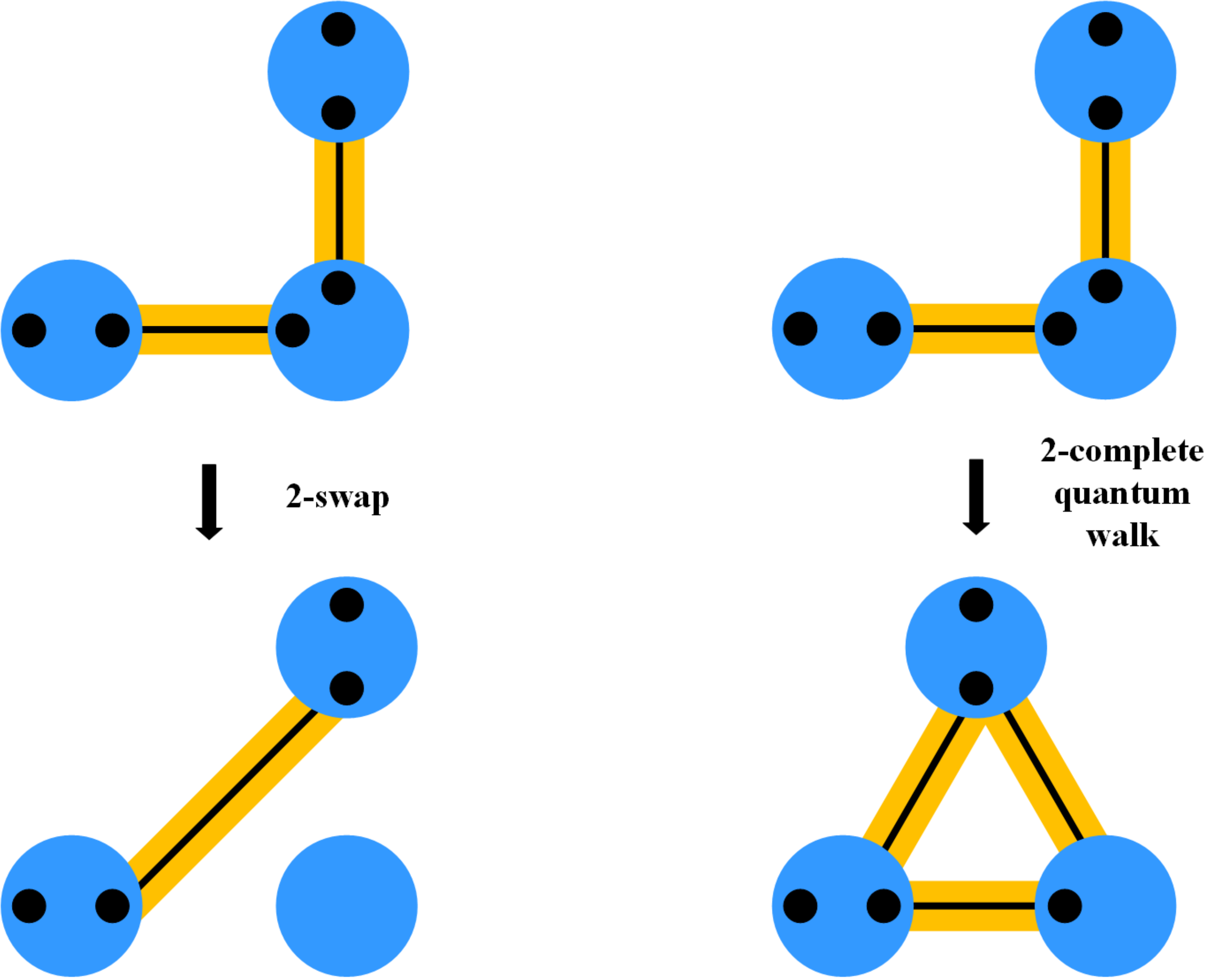}}
\caption{Different structures after 2-swap and 2-complete quantum walk.}
\label{fig13}
\end{figure}

The system is initially in state
\begin{equation}
	\left| \Psi(0)\right\rangle =  \frac{1}{2}(\left| 00\right\rangle + \left| 11\right\rangle)_{12} (\left| 00\right\rangle + \left| 11\right\rangle)_{34}.
	\label{eq7}
\end{equation} 
We take qubit 1 as walker and qubit 2, 3, and 4 as coins and carry out quantum walk in a 2-complete graph. With $C_1=I$ and $C_2=I$, we can generate entanglement between qubit 1, 3, and 4, which is a GHZ state. The transitions are described as
\begin{equation}
	\begin{aligned}
	\left| \Psi(1)\right\rangle = & \frac{1}{2}(\left| 00\right\rangle + \left| 01\right\rangle)_{12} (\left| 00\right\rangle + \left| 11\right\rangle)_{34}, \\
	\left| \Psi(2)\right\rangle = & \frac{1}{2} (\left| 000\right\rangle + \left| 111\right\rangle)_{134} (\left| 0\right\rangle + \left| 1\right\rangle)_{2}. \\
	\end{aligned}
	\label{eq8}
\end{equation}

We can also generalize this to a system with a more common initial state
\begin{equation}
	\left| \Psi'(0)\right\rangle =  (a \left| 00\right\rangle + b \left| 11\right\rangle)_{12} (a \left| 00\right\rangle + b \left| 11\right\rangle)_{34}.
	\label{eq9}
\end{equation}
And we can also achieve entanglement between qubit 1,3, and 4 in a similar way so that 2 partially entangled states can generate entanglement between 3 qubits as well.		
\begin{equation}
	\left| \Psi'(2)\right\rangle =  (a \left| 000\right\rangle + b \left| 111\right\rangle)_{123} (a \left| 0\right\rangle + b \left| 1\right\rangle)_{4}.
	\label{eq10}
\end{equation}
The quantum walk way of pre-processing is shown in Fig.~\ref{fig13}. Compared to q-swap scheme, we can see that different structure with more connections can be formed after 2-complete quantum walk.

\begin{figure}[htbp]
\centerline{\includegraphics[width=0.8\linewidth]{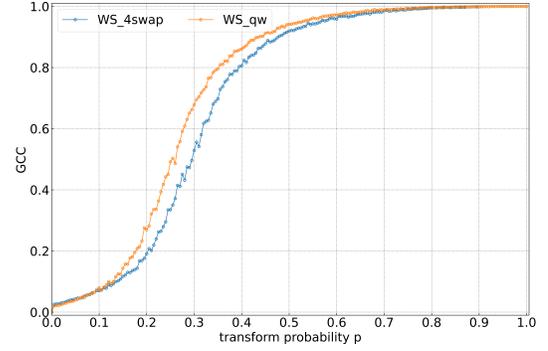}}
\caption{Compare entanglement percolation results of q-swap scheme and quantum walk scheme in a WS network. The 4-swap scheme is indicated by orange line and the quantum walk scheme is indicated by blue line.}
\label{fig14}
\end{figure}

We compare the effects of q-swap and quantum walk in a WS network with nodes of 1000 and average degree of 4. As shown in Fig.~\ref{fig14}, using quantum walk in a 2-complete network can significantly decrease the percolation threshold compared to normal 4-swap scheme. Different from 4-swap scheme, the original connections won’t get destroyed when changing the structure of the network in quantum walk scheme. Therefore, more connections are formed after preprocessing and it takes less time for GCC to reach a similar scale of the whole network. Besides, only local measurements are required in the new scheme, which reduce the error caused by Bell measurements. This makes quantum walk a considerably useful scheme in entanglement percolation.

\section{Conclusions}\label{Section5}
In this paper, we discuss on using entanglement percolation to establish the communication network in small-world quantum networks. In real world, communication networks usually act as small-world networks, so it is of great importance to analyze the percolation operations in small-world networks. Since the percolation threshold of QEP is much lower than that of CEP, we can significantly make the percolation threshold lower by pre-processing the network, which raises the communication efficiency. Besides, results show that different q-swap will lead to different network structures, which have different percolation thresholds. Although raising can increase the value of degree of nodes, the number of these nodes remains small. When is equal to the average degree, the percolation threshold is minimized. Thus, should be determined based on the average degree of the entire network. Finally, we show that multi-coin quantum walk in 2-complete graph can be used to generate more entanglement connections between nodes, which can reduce the percolation threshold even more.


\vspace{12pt}
\end{document}